\documentclass[11pt]{article}

\usepackage{amssymb,amsfonts,amsmath}
\usepackage{natbib}
\usepackage{graphicx}
\bibpunct{(}{)}{,}{a}{,}{,}
\begin{document}

\title{{\bf Model choice versus model criticism}}

\author{{\sc Christian P.~Robert$^{1,2}$, Kerrie Mengersen$^{3}$, and Carla Chen$^{3}$}\\\\
$^1${Universit\'e Paris Dauphine,} $^{2}$ {CREST-INSEE, Paris, France,}\\
and $^3${Queensland University of Technology, Brisbane, Australia}}

\date{ }

\maketitle

\begin{abstract}
The new perspectives on Bayesian model criticisms presented in \cite{ratmann:andrieu:wiuf:richardson:2009}
are challenging standard approaches to Bayesian model choice.  We discuss here some
issues arising from the approach, including prior influence, model assessment and
criticism, and the meaning of error.


\end{abstract}

In \cite{ratmann:andrieu:wiuf:richardson:2009}, the perception of the approximation error in the
ABC algorithm \citep{pritchard:seielstad:perez:feldman:1999,beaumont:zhang:balding:2002,marjoram:etal:2003}
is radically modified, moving from a computational parameter that is calibrated by the user when balancing 
precision and computing time into a genuine parameter $\epsilon$ about which inferences can be made in the same manner as 
for the original parameter $\theta$. As stressed in Section S2 of \cite{ratmann:andrieu:wiuf:richardson:2009}, this is indeed a
change of perception rather than a modification of the ABC method in that the target in $\theta$ remains the same.
(This should not be construed as a criticism in that the unification of most ABC representations proposed in
Section 2 is immensely valuable.) Although the derivation of the distribution $\xi_{x_0,\theta}(\epsilon)$ is somewhat 
convoluted in Section S1, we note here that it is simply the distribution of the error 
$\rho(\mathbb{S}(x),\mathbb{S}(x_0))$ when $x\sim f(x|\theta)$, i.e.~a projection of $f(x|\theta)$ in probabilistic terms.

\medskip\noindent{\bf Example---}For a Poisson $x_0\sim\mathcal{P}(\theta)$ model, a natural divergence is the
difference $\epsilon=x-x_0$ which is distributed as a translated Poisson $\mathcal{P}(\theta)-x_0$ when conditional
on $x_0$ and which is marginaly distributed as the difference of two iid $\mathcal{P}(\theta)$ variables. 
Since $\epsilon$ thus is an integer valued variable, the supplementary prior $\pi_\epsilon$ should reflect this feature. 
A natural solution is 
$$
\pi_\epsilon(k) \propto 1/(1+k^2)\,,
$$
since the series $\sum_k 1/k^2$ is converging, even though using a proper prior $\pi_\epsilon$ does not appear
to be a necessary condition in \cite{ratmann:andrieu:wiuf:richardson:2009}.\hfill$\blacktriangleleft$

\medskip
The change of perception in \cite{ratmann:andrieu:wiuf:richardson:2009} is based on the underlying assumption
that the data is informative about the error term $\epsilon$, which is not necessarily the case, as shown by the 
previous and following examples.

\medskip\noindent{\bf Example---}For a location family, $x_0\sim f(x-\theta)$, if we take $\epsilon=x-x_0$,
the posterior distribution of $\epsilon$ is
$$
\pi_\epsilon(\epsilon|x_0) \propto \int f(\epsilon+x_0-\theta)\pi_\theta(\theta)\pi_\epsilon(\epsilon)\,\text{d}\theta\,\pi(\epsilon)
$$
and therefore a mostly flat prior $\pi_\theta(\theta)$ with a large support
produces a posterior $\pi_\epsilon(\epsilon|x_0)$ identical 
to $\pi_\epsilon(\epsilon)$ for most values of $x_0$. Conversely, a highly concentrated
prior $\pi_\epsilon(\epsilon)$ hardly modifies the posterior $\pi(\theta|x_0)$.\hfill$\blacktriangleleft$

\medskip\noindent{\bf Example---}For the binomial model $x_0\sim\mathcal{B}(n,\theta)$, assuming a
uniform prior $\theta\sim\mathcal{U}(0,1)$, we can consider $\epsilon=x-x_0$, in which case $\epsilon$
is supported on $\{-n,\ldots,n\}$. If we use a uniform prior on $\epsilon$ as well,
\begin{eqnarray*}
\pi_\epsilon(\epsilon|x_0) &\propto& {n \choose \epsilon+x_0} \int \theta^{\epsilon+x_0}
(1-\theta)^{n-\epsilon-x_0} \,\text{d}\theta\,\mathbb{I}_{\{-n,\ldots,n\}}(\epsilon) \\
&\propto& {n \choose \epsilon+x_0} \dfrac{(\epsilon+x_0)!(n-\epsilon-x_0)!}{(n+1)!}\mathbb{I}_{\{-n,\ldots,n\}}(\epsilon)\\
&=& 1\big/(1+2n)\,\mathbb{I}_{\{-n,\ldots,n\}}(\epsilon)
\end{eqnarray*}
and therefore the (Bayesian) model brings no information about $\epsilon$.\hfill$\blacktriangleleft$

Obviously, this example is not directly incriminating against the method of 
\cite{ratmann:andrieu:wiuf:richardson:2009},  in that it only considers a single statistic, instead
of several as in \cite{ratmann:andrieu:wiuf:richardson:2009} (which distinguishes this paper from the
remainder of the literature, where $\epsilon$ is a single number).

\section{Bayesian model assessment}
The paper chooses to assess the validity of the model based on the marginal likelihood $m(x)$ instead of
the predictive $p(x|x_0)$. While this has the advantage of ``using the data once", it suffers from a strong
impact of the prior modelling and of not conditioning on the observed data $x_0$. A more appropriate (if still ad-hoc)
procedure is to relate the observed statistics $\mathbb{S}(x_0)$ with statistics simulated from $p(x|x_0)$, as in, e.g.,
\cite{verdinelli:wasserman:1998}. It may be argued that checking the prior adequacy is a good thing, but having no way to
distinguish between prior and sampling model inadequacy is a difficulty, as seen in the Poisson example.

\medskip\noindent{\bf Example---}For the location family, $x_0\sim f(x-\theta)$,
the joint posterior distribution of $(\theta,\epsilon)$ is
$$
f(\epsilon+x_0-\theta)\pi(\theta)\pi(\epsilon)\,,
$$
and therefore the difference $(\epsilon-\theta)$ is not identifiable from the data, solely from the prior(s).\hfill$\blacktriangleleft$

\medskip
Note that, from an ABC perspective, using $p(x|x_0)$ instead of $m(x)$ does not 
imply a considerable increase in computing time. However, computing the Bayes factor
(and therefore the evidence) using the acceptance rate of the ABC algorithm is even faster.
Moreover, it provides a different answer.

\begin{figure}[h]
\centerline{\includegraphics[width=.4\textwidth]{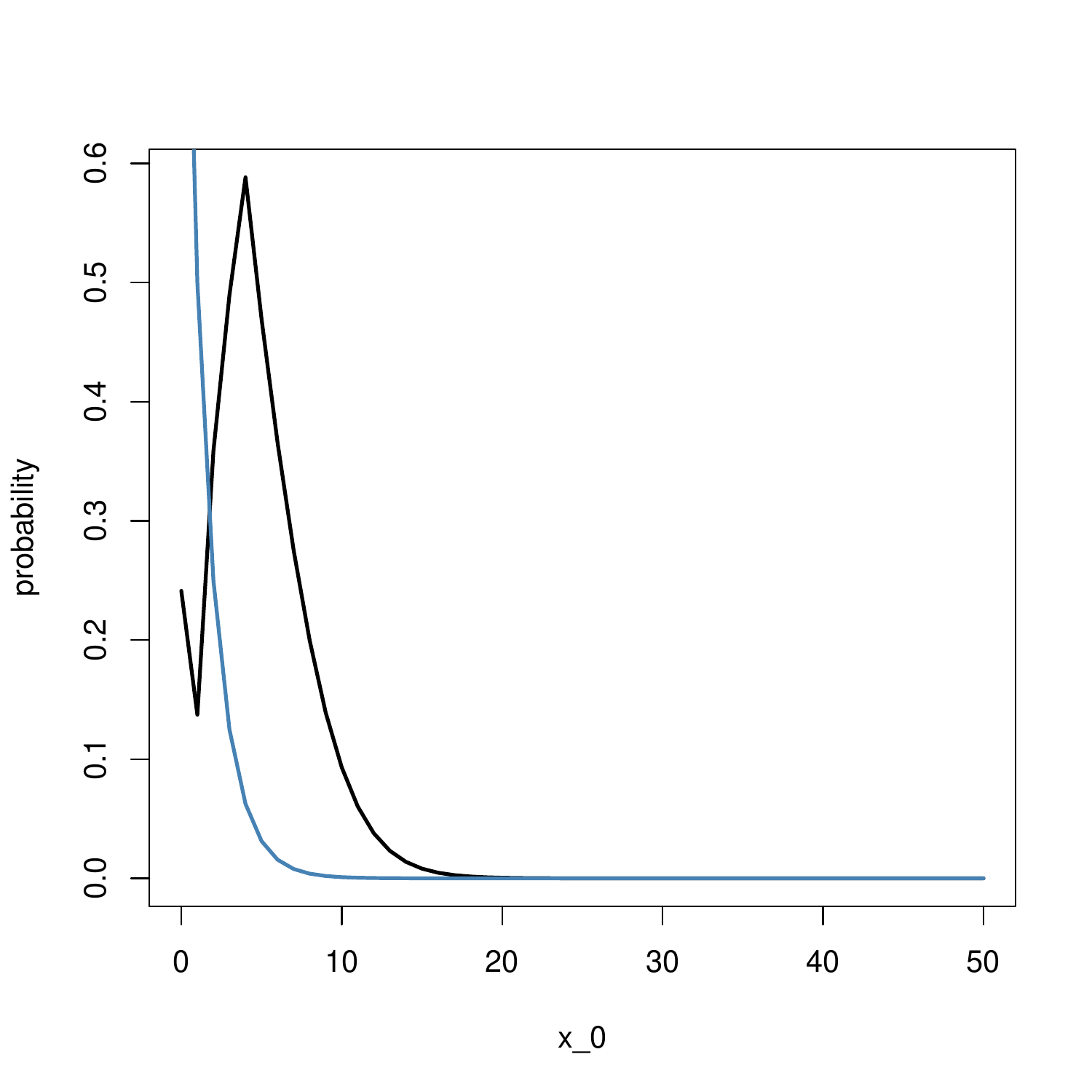}}
\caption{\label{fig:pval}
Comparison of the decreasing rates of the evidence {\em (blue)} and of the $p$-value {\em (black)} 
derived from \cite{ratmann:andrieu:wiuf:richardson:2009} for a Poisson model.}
\end{figure}

\medskip\noindent{\bf Example---}For the Poisson $\mathcal{P}(\theta)$ model, if we take as an example
an exponential $\mathcal{E}(1)$ prior $\pi_\theta$, the evidence associated with the model is 
$$
\int \pi_\theta(\theta) f(x_0|\theta) \text{d}\theta = \int \frac{\theta^{x_0} e^{-2\theta}}{x_0!} \text{d}\theta
	      = 2^{-x_0-1}\,,
$$
while the quantitative assessment of \cite{ratmann:andrieu:wiuf:richardson:2009} is
\begin{equation}\label{eq:pvalue}
\sum_{k=-x_0}^\infty \pi_\epsilon(k|x_0) \mathbb{I}\left\{\pi_\epsilon(k|x_0)\le\pi_\epsilon(0|x_0)\right\}\,,
\end{equation}
with
$$
\pi_\epsilon(\epsilon|x_0) \propto \int \frac{\theta^{\epsilon+x_0} e^{-2\theta}}{(\epsilon+x_0)!(1+\epsilon^2)} \text{d}\theta
= \frac{2^{-\epsilon-x_0-1}}{(1+\epsilon^2)}\,.
$$
The numerical comparison of both functions of $x_0$ in Figure \ref{fig:pval}
shows a much slower decrease in $x_0$ for the $p$-value \eqref{eq:pvalue} than for the evidence,
not to mention a frankly puzzling non-monotonicity of the $p$-value. \hfill$\blacktriangleleft$

\section{Implications of model criticism}
While the approach by \cite{ratmann:andrieu:wiuf:richardson:2009} provides an informal assessment that can be
derived in an ABC setting, the Bayesian foundations of the method may be questioned. The core
of the Bayesian approach is to incorporate all aspects of uncertainty and all aspects of decision consequences
into a single inferential machine that provides the ``optimal" solution. In the current case, while the consequences of
rejecting the current model are not discussed, they would most likely include the construction of another model.
In the first graph in the paper, several models are contrasted and this leads us to wonder about the gain compared with
using the Bayes factor, which can be directly derived from the ABC simulation as well since the (accepted or 
rejected) proposed values are simulated from $\pi(\theta)f(x|\theta)$. 

\medskip\noindent{\bf Example---}For the Poisson $x_0\sim\mathcal{P}(\theta)$ model, running ABC with no approximation
(since this is a finite setting) produces an exact evaluation of the evidence.\hfill$\blacktriangleleft$

\medskip
We also note that the non-parametric evaluation at the basis of the $\text{ABC}_\mu$ algorithm 
of \cite{ratmann:andrieu:wiuf:richardson:2009} can equally be used for approximating the true
marginal density $m(x)$. The smooth version of $\text{ABC}_\mu$ presented in Section S1.5, eqn.~[S8],
is however far from being a density estimate of $\xi_{x_0,\theta}(\epsilon)$ since it based on a single
realisation from $f(x|\theta)$. It should rather be construed as a (further) smoothed version of its smooth
ABC counterpart and this suggests integreting over $h$ as well. Unless some group structure can be exploited
to avoid the repetition of simulations $x_b=x_b(\theta)$, the non-parametric estimator [S9] cannot be used as
a practical device because either $B$ is small, in which case the non-parametric approximation is poor, or $B$
is large, in which case producing the $x_b$'s for every value of $\theta$ is too time-consuming. Obviously, using
moderate $B$ is always feasible from a computational point of view and it can also be argued that the approximation
of $f_\rho(\theta,\epsilon|x_0)$ by $\hat f_\rho(\theta,\epsilon|x_0)$ is not of major interest, since the former is
only an approximation to the true target. (In a vaguely connected way, the rejection sampler of Subsection S1.8 does
seem an approximation to exact rejection-sampling, in that the choice of the upper bound 
$C=\max_i\min_k \hat\xi_k(\epsilon_{ik},\mathbf{x}_i)$ over the samples simulated in Step 1 of the
algorithm does not produce a true upper bound.)

\section{On the meaning of the error}
The error term $\epsilon$ is defined as part of the model, based on the marginal, with the additional input of a
prior distribution $\pi(\epsilon)$. Since \cite{ratmann:andrieu:wiuf:richardson:2009} analyse this error based on
the product of two densities, $\xi_{x_0,\theta}(\epsilon)\pi(\epsilon)$, this product is not properly defined 
from a probabilistic point of view. The authors choose to call $\xi_{x_0,\theta}(\epsilon)$ a ``likelihood" by
a fiducial argument, but this is (strictly speaking) not [proportional to] a density in $x_0$. Obviously, simulating
from the density that is proportional to $\xi_{x_0,\theta}(\epsilon)\pi(\epsilon)\pi(\theta)$ is entirely possible as
long as this function integrates in $(\theta,\epsilon)$ against the dominating measure, but it suffers from an
undefined probabilistic background in that, for instance, it is not invariant under reparameterisation in $\epsilon$:
changing $\epsilon$ to $\varepsilon$ introduces the squared Jacobian $|d\epsilon/d\varepsilon|^2$ in the ``density".
We acknowledge that most ABC strategies can be seen as using a formal ``prior+likelihood" representation of 
the distribution of $\epsilon$, since
$$
\pi_{\text{ABC}}(\theta) = \int \pi_\epsilon(\epsilon)\xi(\epsilon|x_0,\theta)\, \text{d}\epsilon\,\pi(\theta)\,,
$$
but this formal perspective does not turn $\epsilon$ into a ``true" parameter and $\pi_\epsilon$ into its prior.
For instance, non parametric $\pi_\epsilon$'s may be based on the observations or on additional simulations.

The denomination of ``likelihood" is thus debatable in that $\xi_{x_0,\theta}(\epsilon)\pi(\epsilon)\pi(\theta)$
cannot always be turned into a density on $x_0$ (or even on a statistic $\mathbb{S}(x_0)$). 

\medskip\noindent{\bf Example---}For the Poisson $x_0\sim\mathcal{P}(\theta)$ model, $\xi_{x_0,\theta}(\epsilon)$
is the translated Poisson distribution $\mathcal{P}(\theta)-\epsilon$, truncated to positive values. While this is
indeed a distribution on $x_0$, conditional on $(\theta,\epsilon)$, it cannot be used as the original Poisson 
distribution, because of the unidentifiability of $\epsilon$.\hfill$\blacktriangleleft$

\medskip
We also think that comparing models via the (``posterior") distributions of the errors $\epsilon$ does not provide
a coherent setup in that this approach does not incorporate the model complexity penalisation that is at the heart
of the Bayesian model comparison tools like the Bayes factor. First, a more complex (e.g., with more parameters) model
will most likely have a more dispersed distribution on $\epsilon$. Second, returning to the first argument of that nore,
the choice of the prior $\pi(\epsilon)$ (and of the error $\epsilon$ itself) is model dependent (as stressed in the paper via the 
notation $\pi(\epsilon,M)$) and the comparison thus reflects possibly mostly the prior modelling instead of the data assessment, 
as shown, again, by the location parameter example. Using the same band of rejection for all models as in Figure 1
of \cite{ratmann:andrieu:wiuf:richardson:2009} thus does not seem possible nor recommendable on a general basis.

\section*{Acknowledgments}
This work was partially supported by the Agence Nationale de la Recherche (ANR, 212,
rue de Bercy 75012 Paris) through the 2005 project ANR-05-BLAN-0196-01 {\sf Misgepop}
and the 2009 project ANR-08-BLAN-0218 {\sf Big'MC} (for C.P.R.). We are grateful to Oliver Ratmann for clarifying several points about
his paper.

\end{document}